\documentclass{osa-article}

\journal{ome}
\articletype{Review Article}
\usepackage{amssymb}
\usepackage{lineno}
\usepackage{bm}
\usepackage{multirow}

\begin{document}

%\title{Time-Varying Photonic Materials: Physical Considerations and Limitations}
%\title{Time-Varying Photonic Materials and Time Crystals: Dispersion and Energy Constraints}
\title{$\hbar \omega$ versus $\hbar \boldsymbol{k}$: Dispersion and Energy Constraints on Time-Varying Photonic Materials and Time Crystals}

\author{Zeki Hayran,\authormark{1} Jacob B. Khurgin,\authormark{2} and Francesco Monticone\authormark{1,*}}

\address{\authormark{1}School of Electrical and Computer Engineering, Cornell University, Ithaca, New York 14853, USA\\
\authormark{2}Department of Electrical and Computer Engineering, Johns Hopkins University, Baltimore, Maryland 21208, USA\\}

\email{\authormark{*}francesco.monticone@cornell.edu} %% email address is required



\begin{abstract}
Photonic time-varying systems have attracted significant attention owing to their rich physics and potential opportunities for new and enhanced functionalities. %to overcome certain limitations of conventional photonic systems. , such as reciprocity and delay-bandwidth constraints. In this context, the duality of space and time in wave physics has led to the investigation of many interesting physical effects in the temporal domain, 
In this context, the duality of space and time in wave physics has been particularly fruitful to uncover interesting physical effects in the temporal domain, such as reflection/refraction at temporal interfaces and momentum-bandgaps in time crystals. However, the characteristics of the temporal/frequency dimension, particularly its relation to causality and energy conservation ($\hbar \omega$ is energy, whereas $\hbar \boldsymbol{k}$ is momentum), create challenges and constraints that are unique to time-varying systems and are not present in their spatially varying counterparts. Here, we overview two key physical aspects of time-varying photonics that have only received marginal attention so far, namely temporal dispersion and external power requirements, and explore their implications. % for the large and fast time-modulations that are necessary, for instance, for photonic time crystals. 
We discuss how temporal dispersion, an inherent property of any causal material, makes the fields evolve continuously at sharp temporal interfaces and may limit the strength of fast temporal modulations and of various resulting effects. %ultimately limiting the maximum modulation frequency the fields can experience.
Furthermore, we show that changing the refractive index in time always involves large amounts of energy. We derive power requirements to observe a time-crystal response in one of the most popular material platforms in time-varying photonics, i.e., transparent conducting oxides, and we argue that these effects are almost always obscured by less exotic nonlinear phenomena. These observations and findings shed light on the physics and constraints of time-varying photonics, and may guide the design and implementation of future time-modulated photonic systems.
%
%point out that for a time-dispersive time-modulated materials an increase of the modulation speed can come at the expense of a lower modulation strength even if the modulated parameter range is kept fixed. Furthermore, due to external power requirements and the obscuring of the relevant dynamic effects, we argue that currently the most viable route to experimentally realize systems that require large and fast temporal modulations, such as photonic time crystals, is to use the second-order nonlinear process in transparent conductive oxides. %It is hoped that these remarks will provide guidance into the design and implementation of photonic time-varying systems and help unveil novel opportunities related to temporal scattering engineering.
%
%It is currently unclear, however, whether the fast modulation speeds with large modulation strengths necessary, for instance, for photonic time crystals, are practically feasible. 
%
%However, while analogies are often drawn between space-varying and time-varying systems (as for ``time crystals'' and ``time lenses''), modulation in time inherently involves transfer of energy ($\hbar \omega$ is energy, whereas $\hbar k$ is momentum), leading to some strict energy trade-offs that will be discussed later in this paper
%
\end{abstract}

%%%%%%%%%%%%%%%%%%%%%%%%%%  body  %%%%%%%%%%%%%%%%%%%%%%%%%%
\section{Introduction}
Although its roots can be traced back to several decades ago \cite{morgenthaler1958velocity, simon1960action, cassedy1963dispersion, cassedy1965waves}, the study of time-varying photonic materials has recently emerged as one of the most active areas of optics and photonics, potentially opening intriguing directions for new physics and applications \cite{sounas2017non, caloz2019spacetime, caloz2019spacetime2, taravati2020space, engheta2021metamaterials, galiffi2022photonics, yin2022floquet}. Just as spatially-varying artificial materials and structures, such as metamaterials, metasurfaces, and photonic crystals, have enriched the photonic design toolbox to gain better control over the light flow and have led to the unveiling of many unexplored physical phenomena \cite{engheta2006metamaterials}, time-varying engineered materials are now being explored by many research groups worldwide as the next stepping stone to achieve a better level of control of wave phenomena \cite{engheta2021metamaterials}. For instance, space-time-modulated structures can realize nonreciprocal effects without the need for magneto-optical materials \cite{yu2009complete,sounas2017non}, which has been one of the main motivating factors for the recent surge of interest in this field. Moreover, a variety of other interesting effects and concepts based on time-varying systems have been reported in recent years, such as antireflection temporal coatings \cite{pacheco2020antireflection}, spacetime cloaking \cite{mccall2010spacetime}, inverse prisms \cite{akbarzadeh2018inverse}, nonreciprocal amplification \cite{galiffi2019broadband}, static-to-dynamic field conversion \cite{mencagli2022static}, temporal Wood anomalies \cite{galiffi2020wood}, correlated-photon-pair generation \cite{cirone1997photon, sloan2021casimir}, Fresnel drag \cite{huidobro2019fresnel}, spectral causality \cite{hayran2021spectral}, topological phase transitions \cite{ozawa2019topological}, among many others \cite{caloz2019spacetime2}. This short list provides a glimpse into the rich physics that can be accessed by bringing the temporal dimension into play. Importantly, many of these works are based on the concept of \emph{space-time duality} \cite{plansinis2017applications}, namely, the duality of space and time in the wave equation, which provides a conceptual foundation for time-varying systems and facilitates the discovery of new effects and potential applications based on time-varying photonics. However, while analogies between space-varying and time-varying systems are certainly insightful, the two domains have certain distinct characteristics that make modulations in time significantly more challenging to physically realize, especially if large and fast temporal modulations are required, as further discussed in the following.

%However, while analogies are often drawn between space-varying and time-varying systems, modulation in time inherently involves transfer of energy ($\hbar \omega$ is energy, whereas $\hbar k$ is momentum), leading to some strict energy trade-offs that will be discussed later in this paper.

Some of the proposed applications of time-varying photonics do not necessarily require ultra-fast temporal variations and/or large modulation strengths, thus their experimental realization is arguably more feasible. This is true, for instance, for optical modulators and optical isolators based on traveling-wave modulations, where the frequency of the temporal perturbation can be significantly smaller than the operational frequency and the refractive index change can be low \cite{yu2009complete, lira2012electrically, phare2015graphene, li2020lithium} %\red{(with the caveat that, for spacetime-modulated optical isolators, the length $L$ of the device tradeoffs with modulation frequency $\Omega$: if perfect phase matching between two guided modes is guaranteed in one direction, the mismatch in the opposite direction is $\Delta k = 2 \Omega /v_g$ where $v_g$ is the spacetime modulation velocity; then, to achieve isolation, the phase mismatch should be a multiple of $\pi$, leading to a minimum length $L=\pi/\Delta k$)}. 
(with the caveat that, for spacetime-modulated optical isolators, the length of the device trade-offs with the coupling strength between modes \cite{yu2009complete} and the modulation frequency \cite{abdelsalam2020linear}). 
Conversely, for other time-varying systems that have recently been the subject of intense theoretical interest, such as photonic time crystals \cite{biancalana2007dynamics, zurita2009reflection, lustig2018topological, sharabi2021disordered, sacha2020photonic} and broadband optical parametric amplifiers based on momentum bandgaps \cite{lee2021parametric}, the modulation frequency is comparable or larger than the operational frequency, and the modulation strength needs to be sufficiently large to observe the relevant phenomena. For instance, the relative width of the momentum gaps in a time crystal is proportional to the relative modulation strength \cite{reyes2015observation,park2021spatiotemporal} and, therefore, the typical relative permittivity change for these systems is around 10\% \cite{dikopoltsev2022light} and can be as high as 200\% \cite{lustig2018topological}. This ultimately raises the question about whether such fast modulation speeds and large modulation strengths can be physically realized with current or future experimental platforms. Indeed, while the originally proposed time crystals in atomic systems \cite{wilczek2012quantum} have attracted considerable experimental efforts based on various platforms \cite{trager2021real, kessler2021observation,zhang2017observation,choi2017observation,taheri2022all}, experimental demonstrations of photonic time crystals with momentum bandgaps have been scarce and limited to radio frequencies \cite{reyes2015observation}. Although there have been recent promising advances in dynamic optical materials -- especially epsilon-near-zero media \cite{hayran2021capturing} such as transparent conducting oxides -- with experiments showing large femtosecond-scale variation in the refractive index \cite{alam2016large, caspani2016enhanced}, we will show that modulating such materials at the required frequencies implies large power consumption and, moreover, the desired effects are often obscured by less exotic nonlinear phenomena.

%experimental studies on photonic time crystal have not gained momentum yet mainly due to the lack of understanding of the limitations and challenges of current photonic time-varying material platforms.

More broadly, in this article, we discuss two key physical aspects of time-varying materials, namely, temporal dispersion and external power requirements, both of which have important implications for the physical realization and observation of photonic time-varying systems. We argue that, despite the often-invoked duality between space and time, the characteristics of the temporal dimension -- particularly its relation to causality and energy conservation -- can bring up challenges that are unique to time-varying systems.

%\begin{figure}[h!]
%\centering\includegraphics[width=10cm]{dynamic_oscillators.eps}
%\caption{(a) \red{In a dispersive material, an optical field does not react instantaneously to an external temporal modulation and, therefore, its analysis requires care in the presence of fast modulations. [what does not react? the material properties are assumed to react instantaneously, it is the optical response that is not instantaneous] } (b) Two simplified illustrations of time-varying dispersive materials where the number of free-carriers, and therefore the plasma frequency, (left) or the resonance frequency (right) is dynamically modulated \red{[again, need to explain what physical mechanism can cause this]}. While, the free-carrier modulation does not significantly affect the dynamics of the carriers that are already in interaction with the probe beam \red{[probe beam illustration?}], a resonance frequency modulation will affect such carrier dynamics leading to marked differences in the response of the material depending on the relative time scales of the modulation and probe wave (adiabatic vs. non-adiabatic regimes).}
%\label{dispersive_modulation}
%\end{figure}

\section{Dispersion Constraints in Time-Varying Materials}

Most of the studies on time-varying photonics so far have assumed a rather simple dynamic material model, where the time-varying material is non-dispersive and can respond to the external modulation instantaneously. The time-domain constitutive relation for the induced polarization that results from such an approach can be written as $\textbf{P}(t)=\epsilon_0\chi(t)\textbf{E}(t)$ \cite{zurita2009reflection, martinez2016temporal}, where $\textbf{E}$ and $\textbf{P}$ are respectively the electric field and the electric polarization density, while $\chi(t)$ is the time-varying, isotropic, non-dispersive electric susceptibility and $\epsilon_0$ is the free-space permittivity. A time-varying non-dispersive loss (and/or gain) profile can then be modeled by a time-varying conductivity $\sigma(t)$ \cite{song2019direction} that relates the electric field to the electric current density $\textbf{J}(t)$, i.e., $\textbf{J}(t)=\sigma(t)\textbf{E}(t)$. Such simplified relations can be physically interpreted as representing fields that exist in a quasi-steady-state at every time instant and that can acquire a new quasi-steady-state instantaneously as a result of the applied modulation. Clearly, this dynamic model is overly simplistic and, while it can provide approximately correct results in certain cases, it requires proper justification in its usage. Since all physical materials posses a causal, non-instantaneous polarization response with a finite temporal duration, the material constitutive parameters are inherently dispersive, i.e., frequency-dependent. In the linear time-invariant case, this translates into the constitutive relation, $\textbf{P}(t)=\epsilon_0 \int_{-\infty}^{\infty}\chi(t-t')\textbf{E}(t')dt'$, which means that $\textbf{P}$ at time $t$ depends on $\textbf{E}$ not only at $t$ but also at previous time instants (causality implies that $\chi(t-t') = 0$ for $t - t' \leq 0$, namely, no polarization is induced before the field is applied). If the material is time-variant, one can generalize this convolution relation using a time-varying susceptibility kernel, i.e.,
\begin{equation}
\label{constitutive}
    \textbf{P}(t)= \epsilon_0\int_{-\infty}^{\infty} \chi(t,t')\textbf{E}(t')dt',
\end{equation}
as was studied several decades ago (for instance in Refs. \cite{pikulin1973kinetic, kravtsov1974geometrical, stepanov1976dielectric, claasen1982stationary}). This formulation has recently received renewed interest and has been further investigated in Refs. \cite{mirmoosa2022dipole, solis2021functional, sloan2021casimir}.

A major consequence of material dispersion is that the time-varying electric susceptibility depends explicitly on two independent time variables, $t$ and $t'$, as in Eq. (\ref{constitutive}), rather than their difference $t-t'$ as in the time-translation invariant case. Importantly, $\chi(t,t')$ is still causal at every $t$-time instant, so that $\chi(t,t') = 0$ for $t - t' \leq 0$, which results in the generalization of the well-known Kramers-Kronig relations (Hilbert transforms) for time-varying systems, as was studied in Ref. \cite{claasen1982stationary} and more recently in Ref. \cite{solis2021functional}). To gain more physical insight, one can resort to standard differential equation model that relate $\textbf{P}$ to $\textbf{E}$. For instance, for the case of a time-varying dispersive isotropic material of Drude-Lorentz type, the relevant differential equation reads
\begin{equation}
    \label{lorentz_differential}
    \frac{\partial^2 \textbf{P}(t)}{\partial t^2} + \gamma(t)\frac{\partial \textbf{P}(t)}{\partial t} + \omega_0^2(t)\textbf{P}(t) = f\epsilon_0 \omega_p^2(t) \textbf{E}(t),
\end{equation}
where $\gamma$, $\omega_0$, $f$ are, respectively, the damping coefficient, the resonance/natural frequency, and the oscillator strength (for a Drude-type dispersive model, $f=1$), while $\omega_p = \sqrt{(N_1 - N_2)e^2/\epsilon_0m^*}$ is the plasma frequency of the material with $m^*$ being the effective carrier mass and $N_1 (N_2)$ the density of atoms/carriers in the lower (excited) level. The resonant frequency $\omega_0$ is typically some average transition between the valence and conduction band, and one can write $\hbar\omega_0 = E_{\mathrm{gap}}$ (which is often called the Penn gap \cite{ravindra1979penn}). An important assumption made in Eq. (\ref{lorentz_differential}) is the low-density approximation, i.e., the electrons interact weakly with each other such that the dynamics of an electron can be assumed to be independent from other electrons, with or without the presence of an external modulator.

If one considers temporal modulation of the above material, both $\omega_0$ and $\omega_p$ can get modulated. The most obvious option is a modulation of $N_1 - N_2$, achieved through absorption of light via all-optical modulation or through carrier injection/depletion in an electro-optical modulator, such as Si \cite{xu2005micrometre}, ITO \cite{sorger2012ultra} or graphene \cite{phare2015graphene} based modulators. For a Drude-type dispersive system, $\omega_p$ can also be changed by simply re-arranging carriers inside a non-parabolic conduction band, leading to an increase of the effective mass as the carrier temperature increases \cite{secondo2020absorptive}. In most of the faster modulators, such as the electro-optic ones based on Pockels \cite{thomaschewski2022pockels} or Kerr \cite{chakraborty2020cryogenic} effects, acousto-optic ones \cite{savage2010acousto}, and many all-optical ones \cite{yu2016all}, both the effective resonant frequency $\omega_0$ and, to a lesser degree, $f \omega_p^2$ change. This happens when the electronic states shift and mix, which changes $\omega_0$ and (to a lesser degree) the oscillator strength $f$. On the microscopic level, this shifting and mixing can be directly caused by the electric or optical field via DC or AC Stark effects \cite{knox1989femtosecond} or, as in the case of acousto-optic and (partially) Pockels effects, the shifting/mixing is mediated by the re-arrangement of the lattice ions. But overall, one can always model the change of dielectric permittivity via these two parameters -- mean resonant frequency and plasma frequency (strength of the transition).

%which may be modulated in time through various external means, including thermo-optical effects, electro-optical effects (current-driven, e.g., free-carrier modulation, or voltage-driven, e.g. the Stark effect), and all-optical mechanisms \cite{shaltout2019spatiotemporal, reshef2019nonlinear, rahm2013thz, yeh2017ultrafast, guo2020parametrically, im2019iterative}.

In general, the electric susceptibility in Eq. (\ref{constitutive}) can be found by determining the causal Green's function $G(t,t')$ of the differential operator on the left side of Eq. (\ref{lorentz_differential}) (assuming again an isotropic medium)
\begin{equation}
    \label{lorentz_green}
    \begin{split}
    [\frac{\partial^2}{\partial t^2} + \gamma(t)\frac{\partial}{\partial t} + \omega_0^2(t)]G(t,t') = \delta(t-t'), \\
    \chi(t,t') = f \omega_p^2(t)G(t,t'),
    \end{split}
\end{equation}
where the time-varying susceptibility may indeed depend on $t$ and $t'$ explicitly and independently, instead of simply the time passed since the excitation, $t-t'$. In the absence of any time modulation, the electric susceptibility can be easily found and, assuming small losses (i.e., $\omega_0 > \frac{\gamma}{2}$), it can be written as
\begin{equation}
    \label{ti_lorentz}
    \chi(t,t') = \chi(t-t') = f(\omega_0, \omega_p, \gamma, t-t') = 2\pi f \omega_p^2 e^{-\frac{\gamma}{2}(t-t')} \frac{\sin{(\sqrt{\omega_0^2 - \frac{\gamma^2}{4}}} (t-t'))}{\sqrt{\omega_0^2 - \frac{\gamma^2}{4}}}
\end{equation}
for $t-t'>0$, whereas $\chi(t,t') = 0$ for $t-t' \leq 0$. On the other hand, if the material parameters are modulated in time, $\chi(t,t')$ may not be easily found except for several simplified cases depending on the time scale and the specific time-modulation profile. For example, a simplified assumption which is already embedded in Eqs. (\ref{lorentz_differential}) and (\ref{lorentz_green}) is the low-density approximation, implying that the movement of each electron is independent of other electrons, as mentioned above. As a result of this approximation, a modulation of the free-carrier density will only affect $\omega_p$ without affecting the Green's function of Eq. (\ref{lorentz_differential}). Hence, the time-varying susceptibility can be simply found by replacing $\omega_p$ with the time-varying $\omega_p(t)$ in the time-invariant susceptibility function, i.e., $\chi(t,t') = f(\omega_0,\omega_p(t),\gamma,t-t')$, as was also recently discussed in Ref. \cite{mirmoosa2022dipole}. This can be intuitively understood from the fact that injecting or removing carriers can be thought of as an external process that does not significantly affect the dynamics of the carriers that are already in interaction with the probe wave, in accordance with the low-density approximation. Hence, the time-varying susceptibility can be approximated by the time-invariant susceptibility scaled at each time instant by the relevant time-varying factor, proportional to the instantaneous carrier density.

\begin{figure}[h!]
	\centering\includegraphics[width=11cm]{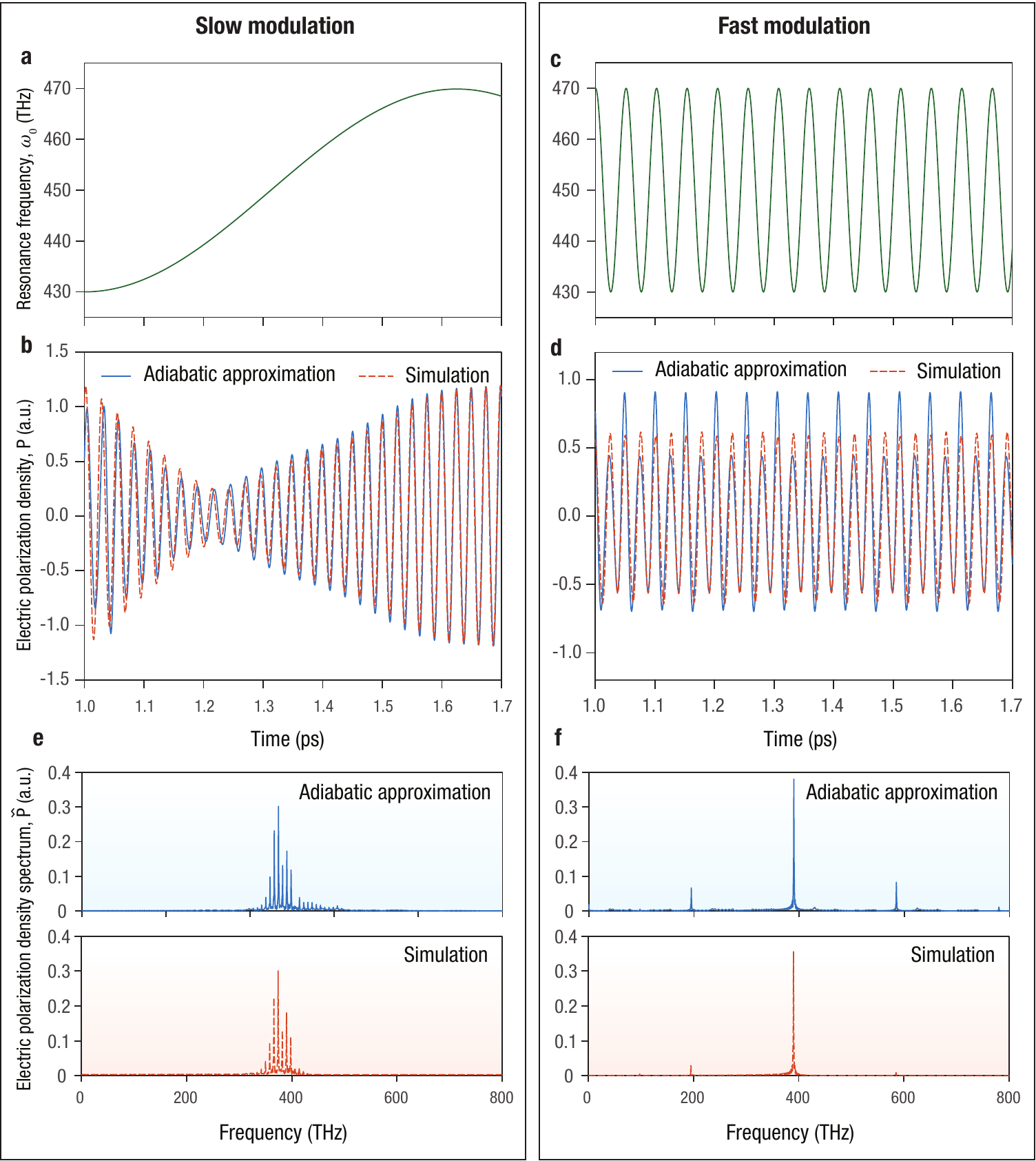}
	\caption{(a) For a slow, adiabatic time modulation, the time scale $\Delta t$ (period) of the modulation is much larger than the time period $\Delta t'$ of the excitation field. (b) As a result, the adiabatic approximation holds and the time-varying dispersive susceptibility is well approximated by the time-invariant dispersive susceptibility function but with time-variant parameters (in this case $\omega_0(t)$). The resulting electric polarization density, calculated with and without adiabatic approximation, is plotted in the figure. (c) If the time scale $\Delta t$ is comparable to $\Delta t'$, (d) the adiabatic approximation becomes inaccurate as the field does not have enough time to interact fully with the dispersive material, and reach a new steady state while the material properties are varied. As a result, the adiabatic approximation may overestimate the strength of the resulting polarization density modulation and of various effects that depend on it, e.g., scattering at temporal interfaces, harmonic generation, and momentum bandgaps in time crystals. (e),(f) Spectra of the polarization density for the slow and fast modulation cases, respectively. Material properties for the two cases are as follows: $\omega_p = 2\pi ~ 400 ~ \textrm{THz}$, $\gamma = 2\pi ~  2 ~\textrm{THz}$, while the excitation signal is monochromatic with a frequency of 390 THz. The modulation frequency is 8 THz and 195 THz for the slow and fast modulation cases, respectively.}
	\label{adiab_vs_nonadiab}
\end{figure}

The situation becomes more involved if an external time modulation affects the dynamics of the carriers that are already in motion due to the probe wave. This may be the case, for instance, if $\omega_0$ is modulated in time (for example using any effect that shifts the relevant energy levels). In this case, $\chi(t,t')$ cannot simply be found, in general, by replacing the time-invariant $\omega_0$ with the time-varying one, and $G(t,t')$ in Eq. (\ref{lorentz_differential}) will need to be evaluated (numerically in most cases). However, if the time scale of the modulation is sufficiently large compared to the period of the probe wave, one can still use the adiabatic assumption to approximate the time-varying susceptibility as $\chi(t,t') \approx f(\omega_0(t),\omega_p,\gamma,t-t')$ \cite{caloz2019spacetime}. In this regard, to emphasize the distinction between the adiabatic and non-adiabatic regimes for time-varying systems, in Fig. \ref{adiab_vs_nonadiab} we provide full-wave simulation results based on the finite-difference time-domain (FDTD) method for slow and fast time-varying systems, using a commercially available software \cite{solutions2021lumerical} and by incorporating the auxiliary differential equation method \cite{kalluri2017advanced, kalluri2018electromagnetics, kalluri2009frequency} to implement the time-varying dispersive material as a custom material plugin (Eq. (\ref{lorentz_differential}) is solved at each time instant to produce the polarization density and displacement field, which then enter Maxwell's equations to update the fields at the next time instant). Specifically, in Fig. \ref{adiab_vs_nonadiab}, we show the polarization density calculated analytically using Eq. (\ref{constitutive}) and the adiabatic approximation for $\chi(t,t')$, compared with the exact full-wave simulation results. It is evident that while the adiabatic approximation leads to accurate results if the time scale of the modulation is large compared to the time scale of the signal wave (and the time scale of the non-instantaneous response of the dispersive material) (Figs. \ref{adiab_vs_nonadiab}(a) and (b)), the approximation starts to break down once the modulation time scale becomes smaller, on the order of the signal wave period (Figs. \ref{adiab_vs_nonadiab}(c) and (d)) since the field does not have enough time to interact fully with the dispersive material and reach a new steady state before the material properties are drastically changed again. As a result, for a fast modulation, an adiabatic approximation may significantly overestimate the strength of the resulting polarization density modulation, as seen in Fig. \ref{adiab_vs_nonadiab}(d), which in turn would lead to an overestimation of the scattering intensity from temporal interfaces and other related effects, such as the size of momentum bandgaps. %If the modulation frequency is further increased, the actual strength of the polarization density modulation would become even lower, until the effect of the modulation vanishes. %As a result, the non-adiabatic regime can suggest a lower susceptibility modulation strength for a given modulated parameter. 

To further elucidate these observations, Figs. \ref{adiab_vs_nonadiab}(e) and (f)) show the spectrum of the polarization density for the slow and fast modulation cases, respectively, obtained through both full-wave simulations and the adiabatic approximation. The amplitudes of the generated harmonics can be used to assess the modulation strength experienced by the propagating wave in the time-varying medium. In the slow, adiabatic case (Fig. \ref{adiab_vs_nonadiab}(e)), the effect of dispersion is weak and the adiabatic approximation is accurate. Conversely, for the fast, non-adiabatic modulation ((Fig. \ref{adiab_vs_nonadiab}(f))), the resulting polarization modulation strength is significantly affected by dispersion, as evidenced by the fact that the adiabatic approximation largely overestimates (by more than 100\%) the intensity of the generated harmonics. %Finally, we have verified that, if the modulation period becomes comparable to the time scale of the non-instantaneous dispersive response of the material, the effects of the modulation completely vanish.

%In the slow, adiabatic case (Fig. \ref{adiab_vs_nonadiab}(e)), since the time-scale of the temporal modulation is large enough compared to the probe wave period, the probe wave can experience the full modulation strength the specific parameter modulation offers [what does it mean?]. However, for the fast, non-adiabatic case ((Fig. \ref{adiab_vs_nonadiab}(e))), the modulation strength becomes weaker as there is not enough time to experience the full parameter modulation range by the probe through the dispersive material \red{[well, you get much stronger harmonic generation at around 200 THz. why? need to explain this]}. Hence, unlike a non-dispersive material (which does not lead to a compromise of the modulation strength as the modulation becomes faster), a dispersive material suggests a trade-off between modulation strength and modulation speed if the operational bandwidth is located near a material resonance (where the material dispersion is more pronounced), which can be ultimately a limiting factor for applications that may require fast temporal modulations together with large modulation strengths. 

\begin{figure}[h!]
\centering\includegraphics[width=12cm]{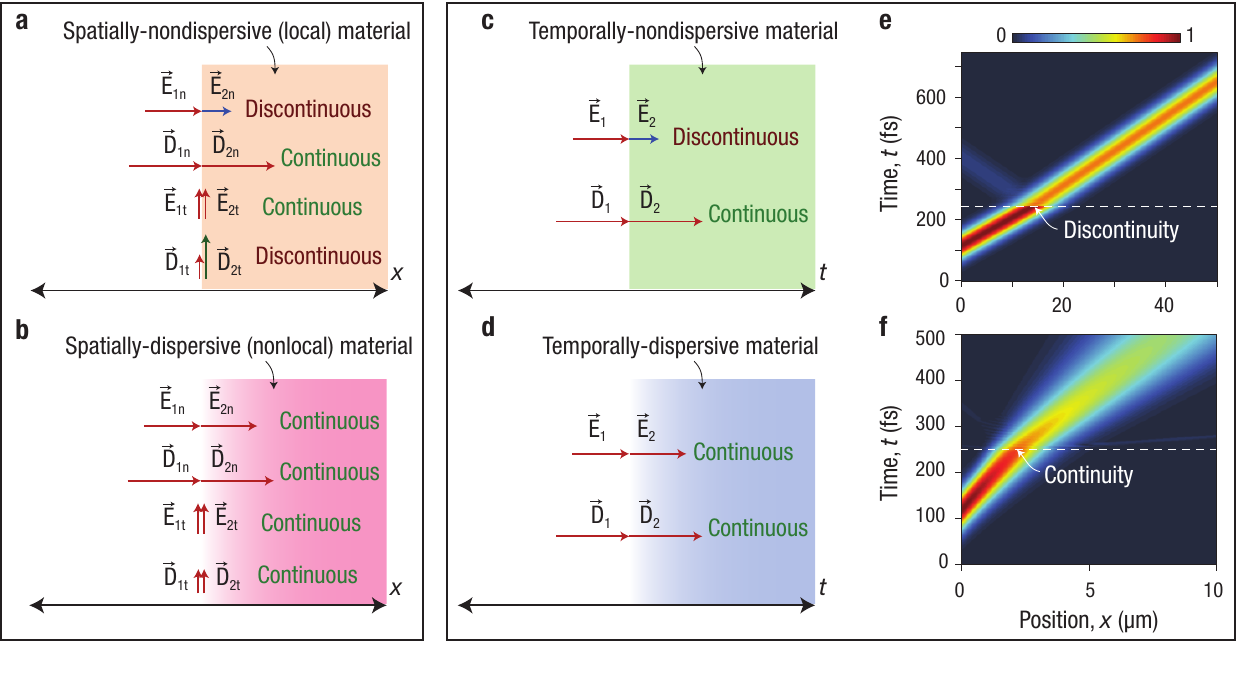}
\caption{(a) At the interface of a spatially-nondispersive (local) material the normal component of the electric field and the tangential component of the displacement field are discontinuous, (b) whereas at the interface of a nonlocal material, with a spatially spread-out material response, these fields should be treated as continuous. (c) Similarly, at the temporal interface of a temporally-nondispersive material, the electric field is discontinuous, (d) whereas at the temporal interface of a temporally-dispersive material, with a temporally spread-out material response, the electric field is continuous due to the non-zero response time of the material. The subscripts "n" and "t" denote the normal and tangential components of the fields, respectively, while the subscripts "1" and "2" denote the fields before and after the spatial/temporal interface, respectively. Full-wave FDTD simulation results (envelope of the electric field) for a time-switched, spatially-homogeneous, (e) nondispersive and (f) dispersive (Lorentz-type) material. Dashed white lines indicate the time-instant when the time-switching of the susceptibility occurs (temporal boundary). For (e), the nondispersive electric susceptibility is switched from 6.42 to 10.05, while for (f) $\omega_p/2\pi$ is switched from 400 THz to 500 THz, with $\omega_0 = 2\pi ~ 430 ~ \textrm{THz}$ and $\gamma = 2\pi ~ 0.02 ~ \textrm{THz}$. In both cases, the incident signal has a bandwidth of 8.8 THz with a central frequency of 400 THz.}
\label{bcs}
\end{figure}

Another important implication of material dispersion arises in relation to the field boundary conditions at a temporal interface (especially in the context of fast temporal modulations, such as time-switching), which has been the subject of debate in the literature \cite{xiao2014reflection, bakunov2014reflection}, and was recently further clarified and reviewed in Refs. \cite{solis2021time,galiffi2022photonics}. Across a spatial boundary the tangential component of \textbf{E} and the normal component of \textbf{D} are continuous, whereas the normal component of \textbf{E} and the tangential component of \textbf{D} are discontinuous (Fig. \ref{bcs}(a))). Interestingly, however, this needs to be revisited in the case of a spatial interface of a spatially-dispersive (nonlocal) material with a spatially extended material response, physically originating from a spread-out charge distribution. This implies that all the fields should be treated as continuous, and gradually changing, across the interface (Fig. \ref{bcs}(b)) \cite{alvarez2020generalized, ciraci2013hydrodynamic}.

On the other hand, across the temporal interface of a time-switched nondispersive material, \textbf{D} is continuous whereas \textbf{E} is discontinuous (Fig. \ref{bcs}(c)) \cite{morgenthaler1958velocity, cirone1997photon, xiao2014reflection}. The continuity of \textbf{D} can be directly proven by integrating Ampere's Law over a vanishing time interval across the temporal boundary. However, similar to the spatially dispersive case, if the time-switched material is temporally dispersive, \textbf{E} will be continuous across the temporal interface as well (Fig. \ref{bcs}(d)) \cite{bakunov2014reflection, solis2021time} due to the non-zero, finite response time of the medium. This can be understood directly from Eq. (\ref{constitutive}) \cite{bakunov2014reflection}: due to the finiteness of $\chi$ and $\textbf{E}$, the resulting $\textbf{P}$ is continuous at the temporal boundary even if the integrand of (\ref{constitutive}) is discontinuous; then, since we already established that $\textbf{D}=\epsilon_0 \textbf{E}+\textbf{P}$ is continuous, $\textbf{E}$ is also continuous. As a result, a propagating wave will effectively experience a relatively gradual change even if time-switching occurs instantaneously. To clarify this point further, in Figs. \ref{bcs}(e) and \ref{bcs}(f), we provide FDTD simulation results (envelope of the electric field) for time-switched non-dispersive and dispersive materials, respectively. It is evident that the electric field experiences a continuous change in its amplitude in the dispersive case even though the material property is switched instantaneously. Hence, special care needs to be taken when determining the boundary conditions in a time-varying system in the presence of dispersion. This is necessary to develop a physically consistent model that is able to accurately predict the behavior of waves at time interfaces, as was already recognized in the literature on rapidly growing plasmas \cite{bakunov2014reflection}. For instance, the results in Figs. \ref{bcs}(e) and \ref{bcs}(f) indicate that using a non-dispersive model for a dispersive material may lead to an overestimation of the reflection from a temporal interface and, therefore, of the width of momentum bandgaps if the temporal variation is periodic. %Moreover, since in dispersive media the field changes gradually and continuously even in the presence of abrupt temporal boundaries, this ultimately limit the maximum speed at which a dispersive material can be meaningfully modulated in time: the $t$-time period of the modulation should be larger than the $t'$-temporal width of the dispersion kernel in Eq. (\ref{constitutive}) , which is on the order of a few tens of fs for plasmonic materials such as transparent conducting oxides. %For additional details on boundary conditions at temporal interfaces, we refer the reader to Ref. \cite{solis2021time}.

\section{Power Constraints in Time-Varying Materials} \label{power}
As mentioned in the introduction, the space-time duality in optics has been a fertile ground in recent years and many of its aspects have been successfully explored. At the same time, while the wave equation looks the same for temporal and spatial variables and there are exact correspondences between the physical phenomena in the two domains, e.g. diffraction and dispersion, there exists one critical difference between perturbations in time and space. The evolution in space domain is characterized by the wave-vector $\boldsymbol{k}$, whereas temporal evolution involves frequency $\omega$. Since $\hbar\boldsymbol{k}$ is a \emph{momentum}, to accomplish change in spatial domain one must supply momentum, which is rather simple -- any stationary perturbation $\Delta n(r)$ of the refractive index (periodic or not) amounts to a momentum change on the scale of $\Delta n ~ \omega/c$. The relative change of index $\Delta n/n$ can be 100\% or more, thus enabling a wide range of modern photonic structures and devices, such as photonic crystals, metasurfaces, etc. What is most important, the momentum is “supplied” without expending energy. In temporal domain, however, $\hbar\omega$ happens to be the \emph{energy}, hence “shaping” light in temporal domain invariably involves a transfer of energy. Specifically, changing the refractive index in time $\Delta n(t)$ always involves prodigious amounts of energy.

One can recognize the scale of the effort required to change the refractive index by a large amount by first noting that the range of refractive indices for all materials in the visible-near IR range is rather small, somewhere between 1.4 and 3.4 \cite{shim2021fundamental}. Low-index materials, like $\mathrm{SiO_2}$, have bandgaps of about 10 eV, while high-index materials, such as GaAs or Si, have bandgaps of about 1 eV. Hence, in order to achieve a large (e.g., 100\%) change of index one must change the bandgap by a few eV. This can be confirmed by using the empirical Moss rule \cite{moss1985relations}, which states that $n^4 E_{\mathrm{gap}} \sim 95$ eV for many materials. Now, changing the bandgap in real time by 1 eV implies changing the bonding energy of each bond by a commensurate amount, and with about few times $10^{23}$ $\mathrm{cm^{-3}}$ valence electrons, the energy required to change the refractive index by 10-20\% is about $10^4 \mathrm{J/cm^3}$ (electric field in excess of $10^8 \mathrm{V/cm}$), i.e., more than enough to cause a breakdown. Furthermore, even if only as little as 0.1\% of the required energy ends up being dissipated inside the material, it would take less than 100 modulation cycles to raise the temperature by 1000 K and melt the material. This suggests that, for relatively large values of $\Delta n/n$, only individual or infrequent changes to the refractive index, as for example in time-switched metamaterials \cite{akbarzadeh2018inverse, pacheco2021temporal, xu2021complete}, may be feasible.

Getting into the specifics, different mechanisms can be conceivably used to change the refractive index, but most of them, such as electro-optic, acousto-optic, let alone thermo-optic, are limited in speed, so no photonic time crystals at optical frequencies can arguably be realized using them, as they require modulation frequencies on the order of the propagating wave frequency (instead, the first two of these mechanisms may be used in applications such as time-modulated optical isolators, which require modulation frequencies significantly smaller than the operational frequency, just enough to couple different modes \cite{yu2009complete}). Neither electro-optic nor acousto-optic modulation can change the refractive index by more than a fraction of a percent, and the speed of electro-optic modulators can reach 100’s of GHz at best \cite{sinatkas2021electro}.

All-optical modulation of the refractive index is the method most actively pursued in this context \cite{pile2016giant}. Confirming our back-of-the-envelope estimates made above, it is well-established that all attempts to change the refractive index via the \emph{instantaneous} third-order optical Kerr effect result in $\Delta n/n$ saturating at less than 1\% at power densities well in excess of 1 TW/$\mathrm{cm^2}$ \cite{borchers2012saturation}. Such power densities are available only in femtosecond laser pulses with low duty cycle, hence one cannot expect to realize photonic time crystals with non-negligible momentum bandgaps (or other time-varying systems with deep modulations) using conventional nonlinear optical materials. 

\begin{figure}[h!]
\centering\includegraphics[width=10cm]{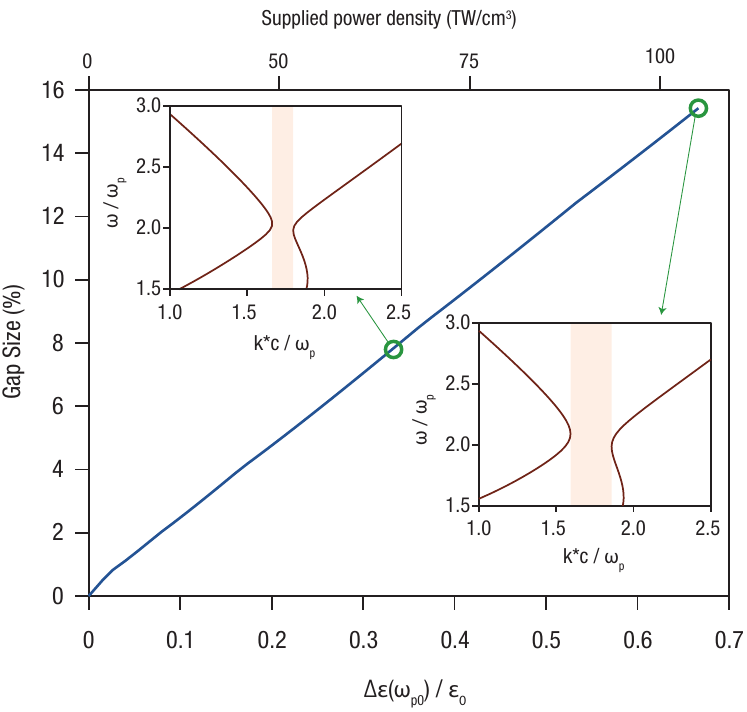}
\caption{Relative momentum-bandgap size (gap/mid-gap ratio) for a frequency-dispersive photonic time crystal, as a function of relative permittivity change (lower horizontal axis) and the corresponding required power density (upper horizontal axis) calculated from Eq. (\ref{energy}) for a modulation frequency $\omega_{\mathrm{mod}} = 10^{13} \mathrm{rad/s}$. A lossless Drude-type dispersive material is assumed and its plasma frequency $\omega_p$ is periodically varied in time as $\omega_p^2 = \omega_{p0}^2 + \Delta\omega_p^2(\cos{(\omega_{\mathrm{mod}}t)}-1)$, where $\omega_{p0} = 1.225\omega_p$, $\Delta\omega_p = 0.707\omega_p$, and $\omega_{\mathrm{mod}} = 4\omega_p$. The insets show the $\omega-k$ dispersion diagram of the photonic time crystal for two values of periodic permittivity modulation, $\Delta\epsilon$, equal to $0.667\epsilon_0$ and $0.333\epsilon_0$ (at frequency $\omega_{p0}$), which both require a supplied power density of more than 50 $\mathrm{TW/cm^3}$ (by assuming a conducting material with $\epsilon_\infty = 1$, $\omega_p = 2.5 \times 10^{12} \mathrm{rad/s}$, $N = 10^{20} \mathrm{cm^{-3}}$, and $E_F = \hbar v_F k_F = 1$ $\mathrm{eV}$). The light-red shaded regions denote the partial momentum bandgaps.}
\label{gap_size}
\end{figure}

Recently, however, different types of nonlinearities have been the subject of great interest -- those in transparent conductive oxides (TCO’s), such as ITO \cite{alam2016large, zhou2020broadband, bohn2021spatiotemporal} and AZO \cite{caspani2016enhanced, kelly2020pump}, where much larger relative changes of the index have been achieved close to the epsilon-near-zero frequency. The dominant nonlinearities in TCOs have a different origin from the familiar Kerr materials, and they are associated with the non-parabolic dispersion of the conduction band and the resulting change of the average effective mass of the electron sea due to intraband absorption \cite{kinsey2019nonlinear, khurgin2021fast, secondo2020absorptive}. Large (on the order of 100\%) changes of refractive index have been indeed observed in these materials, and these changes are fast (few 100’s of fs), albeit still two-to-three orders of magnitude too slow to achieve effects such as time reversal through modulation at twice the signal frequency. Faster nonlinearities are available, but are usually much weaker as discussed below. %and achieve effects such as time reversal  in the optical range. [reference?]} 
One can estimate the power requirements for TCO’s from simple considerations: the change in refractive index or permittivity originates from the change of avarage effective mass, which itself depends on the energy inside the band \cite{secondo2020absorptive}. Therefore, in the most optimistic case, the relative change of the permittivity can be found to be proportional to the relative change $\delta U$ of the energy of carriers: 
\begin{equation} \label{energy}
\delta \varepsilon \sim {\varepsilon _\infty }{{\delta {U_{abs}}} \over {N\hbar {v_F}{k_F}}}, 
\end{equation}
where ${\varepsilon _\infty }$ is the permittivity at high frequencies, assumed on the order of unity, and the average energy of carriers is on the scale of the Fermi energy, i.e., $\hbar {v_F}{k_F} \sim E_\mathrm{f} \sim 1$ eV \cite{secondo2020absorptive}. Since the density $N$ of carriers is few times $10^{20} \mathrm{cm^{-3}}$, this means that to realize a change in permittivity $\delta \varepsilon$ on the order of unity, one has to supply more than 10 J/$\mathrm{cm^3}$ of energy. These considerations provide quantitative insight into the power requirements of photonic time crystals based on TCO-like materials. Fig. \ref{gap_size} shows the numerically calculated momentum-bandgap size for a dispersive photonic time crystal as a function of the required power density that needs to be supplied (note that here the momentum bandgap is not a full bandgap, i.e., it is closed at lower frequencies due to the dispersive nature of the considered material). It is estimated that to obtain a moderately wide momentum bandgap, a power density on the order of tens of TW/$\mathrm{cm^3}$ is needed for THz-scale modulations. Again, this is a very high power, available only in femtosecond lasers, but at least tentatively one may be able to observe photonic time-crystal effects for at least a very short time. The maximum absorbed energy that a material can withstand ultimately limits the maximum possible duration of the periodic temporal modulation and the maximum width of the resulting momentum bandgaps. %In addition, since epsilon-near-zero media are necessarily dispersive to satisfy causality, one should also consider the constraints due to frequency dispersion discussed in the previous section, namely, the modulation period should be larger than the temporal width of the dispersion kernel in Eq. (\ref{constitutive}).

Moreover, when the time crystal is formed all-optically via the third-order susceptibility $\chi^{(3)}$, the time-crystal related effects are almost always obscured by the less exotic phase-conjugation phenomena. Consider the general four-wave mixing process in which two counter-propagating pump fields $E_{\mathrm{p\pm}} = A_{\mathrm{p}} e^{j(\pm\boldsymbol{k}_\mathrm{p}\cdot\boldsymbol{r}-\omega t)}+c.c.$ (where $c.c.$ stands for complex conjugate) interact with the signal wave $E_{\mathrm{s}} = A_\mathrm{s} e^{j(\boldsymbol{k}_\mathrm{s}\cdot\boldsymbol{r}-\omega t)}+c.c.$, and generate a third-order nonlinear polarization $P_\mathrm{NL}$, which includes three terms proportional to $\chi^{(3)}$ due to the mixing of these three waves \cite{shen1984principles}. %that one may write as $P_\mathrm{NL} \sim 3\chi^{(3)}E^3 \sim 3\chi^{(3)}A_\mathrm{p}^2A_\mathrm{s}^*e^{j(-\boldsymbol{k}_\mathrm{s}\cdot\boldsymbol{r}-\omega t)}+c.c.$ \cite{shen1984principles}. 
While at first glance it may appear that all three terms are equal, this is certainly not the case. \emph{The order in which the waves mix is critical.}

The first term is $P_{\mathrm{TR}} \sim \chi^{(3)}(\omega; -\omega, \omega, \omega)A_{\mathrm{p+}}A_{\mathrm{p-}}A_\mathrm{s}^*$ where the subscript "TR" stands for "time reversal". What this physically means is that the two pump waves mix and produce a uniform intensity pattern oscillating at frequency $2\omega$, which modulates the index at this frequency thus forming a time crystal in which Floquet states become possible. The signal wave then scatters off the time crystal into a time-reversed wave. Note that the origin of this “fast” nonlinearity in TCOs is ballistic motion of electrons inside the non-parabolic conduction band and this ultrafast nonlinearity is relatively weak, as discussed in Ref. \cite{khurgin2021fast}. 

The other two terms are $P_{\mathrm{DG}} \sim \chi^{(3)}(\omega; \omega, -\omega, \omega)[A_{\mathrm{p+}}A_{\mathrm{s}}^*A_{\mathrm{p-}} + A_{\mathrm{p-}}A_{\mathrm{s}}^*A_{\mathrm{p+}}]$, where the subscript “DR” stands for “dynamic grating” \cite{shen1986basic}. Here, it is the interference of pump and signal beams that produces the grating. For a CW pump and probe, the grating is stationary, but, since in most experiments short pulses are used, the grating is actually dynamic on the scale of the pulse length, which, compared to optical frequencies, is quite slow. As already mentioned, the origin of this “slow” nonlinearity is an increase in the energy inside the nonparabolic band (increase of electron temperature) due to absorption \cite{khurgin2021fast}, and the strength of this grating is determined by the characteristic response time of the medium. In TCO’s, this response time is the time it takes to cool photoexcited electrons to the lattice temperature $\tau_{\mathrm{el}}$ (typically on the scale of few hundreds of fs) \cite{khurgin2021fast}. The “slow” nonlinearity $\chi^{(3)}(\omega; \omega, -\omega, \omega)$ is much larger than the fast nonlinearity $\chi^{(3)}(\omega; -\omega, \omega, \omega)$, because the index change is integrated over the shorter of pulse length and characteristic time. In particular, in TCO’s the ratio of “slow” to “fast” nonlinearities is on the scale of $2\tau_{\mathrm{el}}/\tau_\mathrm{s}$, where $\tau_\mathrm{s}$ is a momentum scattering time on the scale of a few fs. Hence, the “slow” nonlinearity is about 100 times stronger than fast one. In the more general case, the ratio of the two nonlinearities is $2T_1/T_2$, where $T_1$ is the characteristic lifetime and $T_2$ is the (usually much shorter) coherence time in the nonlinear medium. Therefore, in a medium like TCO and many others the response is always dominated by the relatively slow dynamic grating and not by the time crystal oscillating at $2\omega$. We want to stress that, as far as measurements go, the two responses are externally indistinguishable. One can modify the geometry and use a very thin sub-wavelength layer of nonlinear medium to obtain a forward-propagating “negative refraction” wave in addition to the backward-propagating “phase conjugated” wave \cite{pendry2008time, vezzoli2018optical}, but the response due to the slow dynamic grating will always dominate. The only way to obtain a “pure” time crystal response is to operate near the two-photon absorption edge in transparent materials, such as Si or GaAs, to maximize $\chi^{(3)}(\omega; -\omega, \omega, \omega)$ while avoiding absorption associated with the slow nonlinearity $\chi^{(3)}(\omega; \omega, -\omega, \omega)$, but even then the effect remains quite weak.

It is our belief that one of the most promising paths to observe a time crystal in the optical range is to use not the third but the second order ($\chi^{(2)}$) nonlinear process, in which a standing pump wave at the second harmonic frequency $2\omega$ interacts with counter-propagating waves at the fundamental frequency, which are of course time reversed. This process is instantaneous and is not obscured by spurious dynamic-grating effects. Such an arrangement was first proposed in 1996 \cite{ding1995transversely, ding1996optical} and named ``Transversely Pumped Counterpropagating Optical Parametric Oscillation and Amplification'' (TCOPA). It was demonstrated a decade later \cite{lanco2006semiconductor} when twin counter-propagating photons -- time-reversed replicas of each other -- were produced. The advantage of the TCOPA scheme is that the pump wave is contained inside a resonant cavity which increases the amplitude, $E_{2\omega}$, and thus the effective index change $\Delta n \sim 1/2 ~ \chi^{(2)}E_{2\omega}$. The transverse geometry facilitates effective phase matching in cubic III-V crystals with $\chi^{(2)} > 100$ pm/V, hence one can effectively achieve $\Delta n/n$ of a few percent (and momentum bandgaps of commensurate relative size) with a power density of less than 100 GW/$\mathrm{cm^2}$ inside the cavity, i.e., with only a few GW/$\mathrm{cm^2}$ of incident power.

\section{Conclusions}

In summary, in this article we have discussed various aspects of time-varying photonics that have important implications for the modeling and practical realization of photonic time-varying materials and time crystals. %received relatively less attention in the recent literature of this emerging field.
While the space-time duality in wave physics has provided valuable insight and facilitated the discovery of various new effects, the temporal domain is fundamentally different from the space domain due to its direct relation to the principles of causality and energy conservation: (i) causality implies that any material (except vacuum) has a non-instantaneous, dispersive, electromagnetic response, and (ii) temporal evolution is characterized by frequency $\omega$ and, therefore, shaping light in the temporal domain requires a transfer of energy $\hbar \omega$. %We focused on two key characteristics, i.e., temporal dispersion and energy requirements, important implications for the modeling, operation, and physical realization of photonic time-varying materials and time crystals. 

In the first half of this article, we have reviewed how temporal dispersion is modeled in time-varying systems and how it affects the physical behavior of fields in temporally modulated materials and at temporal interfaces. We showed that, in a dispersive time-varying scenario, a trade-off typically exists between the modulation speed and the strength of the resulting polarization density modulation and, hence, of various effects that depend on it. An adiabatic approximation may lead to largely overestimating these effects. Moreover, in dispersive time-varying materials, all fields are continuous across a temporal interface even when the material properties are modified instantaneously. Then, in the second part of the article, we have discussed the energy requirements to change the refractive index in time based on general considerations and for specific material platforms, focusing on epsilon-near-zero media such as transparent conducting oxides. We found that a modulation sufficiently large and fast to observe a time-crystal response in the optical range requires very high levels of input power, available only in femtosecond lasers, which makes the observation of these effects possible but challenging, and limited to short times.  Moreover, we argued that time-crystal effects based on third-order nonlinearities are almost always obscured by less exotic phase-conjugation phenomena, whereas second-order nonlinearities may offer an interesting path to observe these effects at optical frequencies.

We believe that the rich physics of time-varying systems may open intriguing opportunities for new and advanced functionalities in photonics, which the research efforts so far have provided a valuable glimpse into. However, physical considerations, constraints, and experimental challenges must be carefully assessed before useful applications can emerge from these research efforts.

\begin{backmatter}
\bmsection{Funding}
Air Force Office of Scientific Research (FA9550-22-1-0204) through Dr. Arje Nachman. National Science Foundation (1741694). Office of Naval Research (N00014-22-1-2486).

\bmsection{Acknowledgments}
JK appreciates all the help and encouragement given by his collaborators Prof. P. Noir and Dr. S. Artois.

\bmsection{Disclosures}
The authors declare no conflicts of interest.

\bmsection{Data availability}
Data underlying the results presented in this paper are not publicly available at this time but may be obtained from the authors upon reasonable request.

\end{backmatter}

%%%%%%%%%%%%%%%%%%%%%%% References %%%%%%%%%%%%%%%%%%%%%%%%%
%\bibliographystyle{osajnl}
\bibliography{photonic_time_varying_materials.bbl}

\end{document}